# Toward a quantum concept of absolute simultaneity: Observation of single-electron spatial dynamics with a macroscopic discontinuity

**Sergey A. Emelyanov**

*Ioffe Institute, 194021 St. Petersburg, Russia*

E-mail: sergey.emelyanov@mail.ioffe.ru

**Abstract.** We provide strong evidence for single-electron spatial dynamics with a macroscopic discontinuity. The dynamics is observed in an electron quantum phase consisted of macroscopic quantum orbits similar to those responsible for the integer quantum Hall effect. The dynamics is consistent with the standard QM that predicts a "cloud-like" behavior of electron in quantum orbits regardless of their lengthscale. Since the dynamics is beyond the relativistic paradigm of movement, it may well be nonlocal. It thereby can be the basis for a quantum concept of absolute simultaneity, which now rests not on the meaningless notion of infinite speed but on empirically-tested quantum effect. This concept revives the Bell-Popper idea to replace the current "kinematic" version of relativity by the "dynamic" version by Lorentz and Poincare with a revival of the classical view of space and time where the problem of preferred reference frame now can be solved without to involve the vague notion "aether". We argue that the concept of quantum absolute simultaneity opens the door to a realistic interpretation of standard QM through a deeper insight of what is called "reality" and ultimately to a harmony between quantum and relativity theories.

## Introduction

Despite more than a hundred years of coexistence of relativistic and quantum theories, it has not yet been possible to eliminate conceptual contradictions between them at least in their generally-accepted interpretations. Ultimately, this is related to the fact that the formalism of standard quantum mechanics (QM) addresses neither classical three-dimensional space nor relativistic four-dimensional spacetime but a multi-dimensional Hilbert space which is the space of so-called quantum states. These contradictions concern even the holy of holies of relativism – the spatial dynamics of physical bodies because QM rejects a fundamental description of single-particle spatial dynamics of in terms of trajectories (see e.g. [1]). This means, at least in a general case, quantum spatial dynamics goes beyond the very relativistic paradigm of movement.

In this situation, one would seem that the only possible way to reconcile these theories is to make them disjoint so that each of them has its own application scope. This is exactly what was done when the new-born quantum theory was declared to be concern only the micro-world and moreover the micro-world was declared to be fundamentally unknowable (or even non-existent). In this case, quantum formalism is not more than a mathematical algorithm that fundamentally does not allow any realistic interpretation. Niels Bohr himself expressed this idea in his famous phrase: "*There is no quantum world. There is only a quantum-mechanical description of physical phenomena*".

Today this view of QM remains dominant. Not so long ago, it was clearly expressed by David Mermin [2]. When asked what QM was telling him, he answer in a short phrase "*Shut up and calculate*". And this phrase so vividly reflects the current insight of QM that it immediately becomes a kind of meme among physicists and even something like a philosophical basis for an original view of the entire universe [3]. In fact, however, such a strange (at least for "neophytes") view of QM is not an oddity, but rather a forced step. Indeed, if we assume that QM does describe what Einstein called "objective reality", then, for example, an electron transition between atomic orbits may be regarded as a single-electron discrete spatial dynamics that calls into question such foundation of special relativity (SR) as the suggestion of continuity of any spatial dynamics. But in the frames of Bohr's approach, this problem disappears as here any electron transitions are not more than a comfortable model that has nothing to do with reality.

However, such an "algorithmic" view of QM leads to a serious problem. The point is although QM does deal only with micro-objects, this fact cannot guarantee that quantum laws will never penetrate into the macro-world. At least, the quantum formalism, in itself, has no any fundamental restrictions related to lengthscale. And if such penetration will really take place, then the relativistic picture of physical world may turn to be untenable. In fact, a kind of rehearsal of such a scenario was the discovery of EPR nonlocality. We mean the implementation of the Einstein-Podolsky-Rosen (EPR) thought experiment, when, by using Bell's theorem of 1964, it was shown experimentally that there is a nonlocal correlation in the behavior of entangled photons flying apart [4,5]. The most famous experiments of that kind were carried out by Alain Aspect's group in the early 80's and they caused a wide public outcry even far beyond the physical community [6].

Actually, after the Aspect's experiments, physics appears at a crossroads. On the one hand, the so-called no-communication theorem, which prohibit any signaling in the EPR situation, makes it possible to leave everything "as is" at least as for the question of how to interpret both quantum and relativity theories [7]. This means we preserve both Einstein's SR and Bohr's QM but now quantum theory irrevocably loses a chance to have any realistic interpretation because otherwise we should recognize a direct correlation between the past and the future.

On the other hand, the ultimate rejection of physical realism did not suit those who were the successors of the Einstein's realistic line in the philosophy of science. The most famous among them were John Bell himself and philosopher Karl Popper. They both independently proposed an alternative way of the further development of physics [8,9]. This way rests on the fact that, strictly speaking, the Lorentz-Poincare's "dynamic" version of relativity and the Einstein's "kinematic" version are experimentally indistinguishable. The choice once made in favor of the latter is related only to the fact that the former has an additional problem of preferred reference frame, which seems hard to resolve. Nevertheless, this problem does not look fundamentally unsolvable and, in the case it will be resolved, one would come back to the classical view of space and time where quantum nonlocality has nothing to do with causality.

However, there are serious objections to the Bell-Popper idea. First, the rejection of SR automatically calls into question general relativity (GR) which, in that time, reflected the generally-accepted view of gravity. Secondly, in the minds of most physicists, the "dynamic" version was associated with the problem of "aether" so that it looks much less elegant than the "kinematic" version. Finally, for a long time physicists had an extremely successful experience in using standard QM for all practical purposes. As a result, most of them began to treat the mysticism of QM as something inevitable, and some even saw a deep philosophical meaning in that. Therefore, the very motive for the Bell-Popper idea – to leave a chance for physical realism – did not look convincing to the majority of physical community. Ultimately, all these objections had led to the rejection of the Bell-Popper idea.

At the same time, by the rejection of this idea, physicists become a kind of hostage of a suggestion that has not been proven by anyone and currently is perceived as an axiom. We mean the suggestion that EPR nonlocality is the only realizable quantum nonlocality and accordingly the entanglement is the only quantum phenomenon that gives rise to nonlocality.

In general, this suggestion seems quite reasonable because the penetration of quantum laws into the macro-world is extremely rare indeed. Today, along with the EPR effect, we know only two macroscopic quantum effects. These are superconductivity (and similar effects) and the so-called integer quantum Hall (IQH) effect. But, in the former case, the effect is related to a collective behavior of quantum particles rather than to their individual behavior. In this situation, it is hard to expect any nonlocality. As for the IQH effect, although here a macroscopic quantum effect is truly related to an individual behavior of electrons, it nevertheless is an edging effect where the number of electrons responsible for the effect is always much less than their total number. As a result, such "macroscopic" electrons manifest themselves only in precise magneto-transport measurements and they have never been regarded in the context of an alternative quantum nonlocality.

# Single-electron spatial dynamics through a macroscopic quantum orbit

## *The idea of an alternative nonlocality in IQH-type system*

Leaving aside (for a time) the problem of quantum nonlocality, let us first clarify what are the "macroscopic" electrons that lead to the IQH effect. To this aim, consider what the IQH system is.

As a rule, the source material for IQH system is a two-dimensional semiconductor structure where the thickness of conducting layer is less than the electrons' mean free path in the material (typically, a few tens of nanometers). Accordingly, in two dimensions (*X* and *Y*) we deal with a gas of free electrons of Bloch type whereas in the third dimension (*Z*) the energy spectrum is a series of quantum-size levels. Thus, if a strong magnetic field is applied in the *Z*-direction, then the electron motion becomes quantized in all directions. As a result, each quantum-size level gives rise to a series of sublevels known as Landau levels.

Fig. 1 (left panel) shows the energy spectrum of an idealized IQH system where we neglect the local potential fluctuations. Here *X*-direction is arbitrary because of the axial symmetry in the *XY* plane. Electrons are localized within the so-called cyclotron orbits which are microscopic. Their absolute number in each Landau level is determined by their dense packing over the system in the *XY* plane. The electrons are strongly degenerate with respect to their wave vector in the *Y* direction ($k_y$) or equivalently with respect to the coordinate of the center of their cyclotron orbits in the *X* direction ($X_o$) because these two parameters are in a direct correlation: $X_o = -\lambda^2 k_y$, where $\lambda$ is the so-called magnetic length.

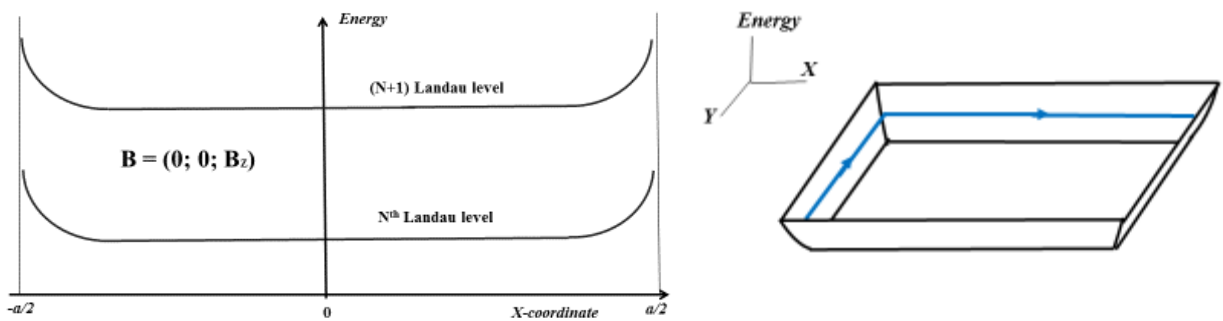

**Fig. 1**. Left panel: Energy spectrum of an idealized IQH system. Local potential fluctuations are neglected. The coordinate scale is strongly exaggerated near the edges to show a microscopic strip with orbit-like quantum states extended along the perimeter of the entire system.
Right panel: The energy diagram of Landau level in two dimensions. One of the current-carrying states is shown in blue.

But this is not the whole picture. We should take into account that in a microscopic proximity to the system's edges which are the boundary between two media, there is always a strong electric field that lies in the *XY* plane and is perpendicular to the edge. As a result, here we get a crossed electric and magnetic field that lifts the Landau level degeneracy and gives rise to spatially-separated orbit-like quantum states along the perimeter of the system. These states with the characteristic cross-section is of the order of the radius of cyclotron orbit were predicted theoretically by Bertrand Halperin who called them extended (or current-carrying) states [10]. These edging states are precisely the reason for the IQH effect, which manifests itself in an exact quantization of transverse conductivity in such system [11]. This discovery by Klaus von Klitzing was awarded the Nobel Prize in 1985 [12].

To clarify the spatial configuration of Halperin's states, we show the energy diagram of a Landau level in two dimensions (Fig. 1, right panel). The diagram looks like a bowl with a large flat bottom (without fluctuations) and extremely narrow walls, on the inner side of which there are the Halperin's states capable of carrying a macroscopic current. Thus, the lengthscale of such orbit-like states has rather technological than fundamental limitations because it is the lengthscale of the macroscopic system itself.

Now, when we have clarified what are the "macroscopic" electrons in the IQH effect, it seems the time to clarify how they could lead to a quantum nonlocality. To this aim, consider a thought experiment, the conditions of which are very close to a real IQH system (Fig. 2).

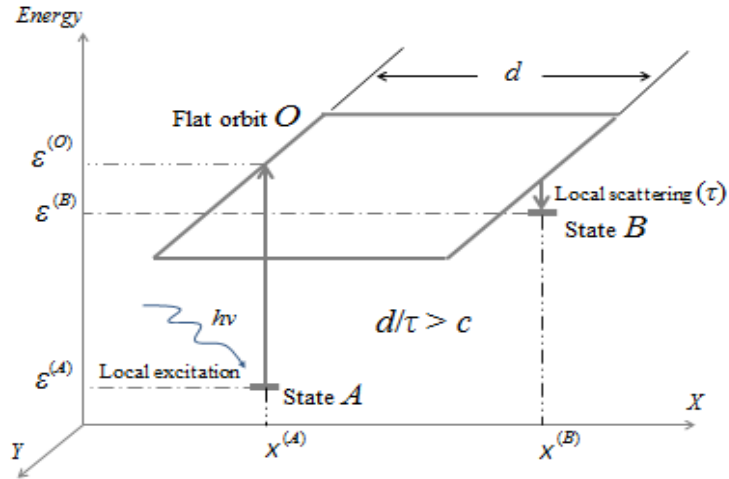

**Fig. 2**. Simplified model of a single-electron discrete spatial dynamics through a flat macroscopic quantum orbit.

Suppose there is a quasi-one-dimensional macro-orbit ($O$) in a two-dimensional electron system with charged point-like scatterers of a high concentration. Since the configuration of the orbit is determined by that of the system edges, suppose that the orbit is a square with a side $d$. Let the electron lifetime in the orbit ($\tau$) is determined by the scattering that transfers electron from the orbit to a nearby local vacant state. Now if a light quantum excites electron into the orbit from a low-energy local state ($A$) close to the left side of the square, then, after the time $\tau$, the electron will undergo a local scattering at any point of the orbit (say, at the point $B$) and occupy a local vacant state close to this point.

But the point is, according to the principles of QM, the electron fundamentally has no any definable position in the orbit or equivalently behaves like a cloud smeared over the orbit. As a result, the probability to be scattered near the left side of the square is exactly the same as the probability to be scattered near the right side (one-fourth). This means that with the probability of one-fourth, the electron will cover the distance not less than *d without having been to any intermediate point*. Therefore, the distance covered may well be many orders longer than the electron mean free path determined by the scattering. In terms of quantum formalism, it is precisely the effect caused by the appeal of quantum formalism not to the Euclidian space but to a multidimensional Hilbert space which is the space of quantum states where any orbit is an indivisible point regardless of its lengthscale.

Thus, if macroscopic orbits do exist, then QM predicts an electron spatial dynamics with an arbitrary macroscopic discontinuity and this dynamics may well be nonlocal insofar as it is beyond the very relativistic paradigm of movement. Indeed, at a fixed $\tau$, we can always take such $d$ that the ratio $d / \tau$ will be higher than the speed of light. Moreover, if we take a typical lengthscale of the IQH systems (about 1cm), and a typical scattering time (about 3ps), then the ratio $d / \tau$ will be one order higher than the speed of light. But this fact does not contradict relativistic formalism simply because the above ratio has nothing to do with what we call speed.

*The idea of a quantum phase with macroscopic electron orbits*

Of course, our thought experiment is rather speculative and, to talk seriously about the discrete spatial dynamics, we need an experimental confirmation. However, it seems hard to observe such dynamics (if any) in conventional IQH system because, as we noted, here the macroscopic states are concentrated near the edges and therefore are detectable only in precise magnetotransport measurements because all the other quantum states cannot contribute into the in-plane macroscopic current.

Nevertheless, there is a chance to avoid this difficulty. The point is that a strong electric field (up to $10^5$ V/cm) may occur not only near the edges but also in the system interior. For example, such a situation arises when the conducting layer is very close to the sample surface. In this case, the so-called surface potential (caused also by the boundary between two media) penetrates into the conducting layer. As a result, there is an asymmetry of confining potential or equivalently an electric field perpendicular to the layer. It is the so-called "built-in" field ($E_{\text{built-in}}$).

Thus, if we take such an asymmetric structure in quantizing magnetic field with a nonzero in-plane component, then crossed electric and magnetic field will occur not only near the edges but also in the system interior. The energy spectrum of a system of that kind was first calculated as long ago as in the early 90's [13-14]. For a simplified model of infinite (in the XY plane) two-dimensional system, it was shown that here the Landau level degeneracy may truly be lifted so that we have got something like energy bands in the X-direction which now is not arbitrary but strictly defined by the in-plane magnetic field ($B_X$). The calculations have shown that, in the frames of this model, spatially-separated single-electron spontaneous currents flow in opposite directions along the Y axis so that the total current along this axis is always zero. Their characteristic cross-section is of the order of the cyclotron radius related to $B_X$ and, at least in this regard, they are reminiscent the Halperin's single-electron orbital currents.

Fig. 3 (left panel) shows the energy spectrum of such system, which now is not strictly symmetric with respect to the vertical axis because there is a small shift of such Landau bands, which increases with increasing of the Landau quantum number (N). Thus, if this spectrum is relevant, then, in a real bounded system, the energy diagram of a Landau level will again be an "energy bowl" but now with a trough-like bottom along the Y axis (Fig. 3, right panel). Accordingly, one would expect that the spontaneous currents of the same energy will be closed near the edges parallel to the X-axis. As a result, we get a system of macroscopic quantum orbits those are no longer concentrated near the edges but cover the whole system.

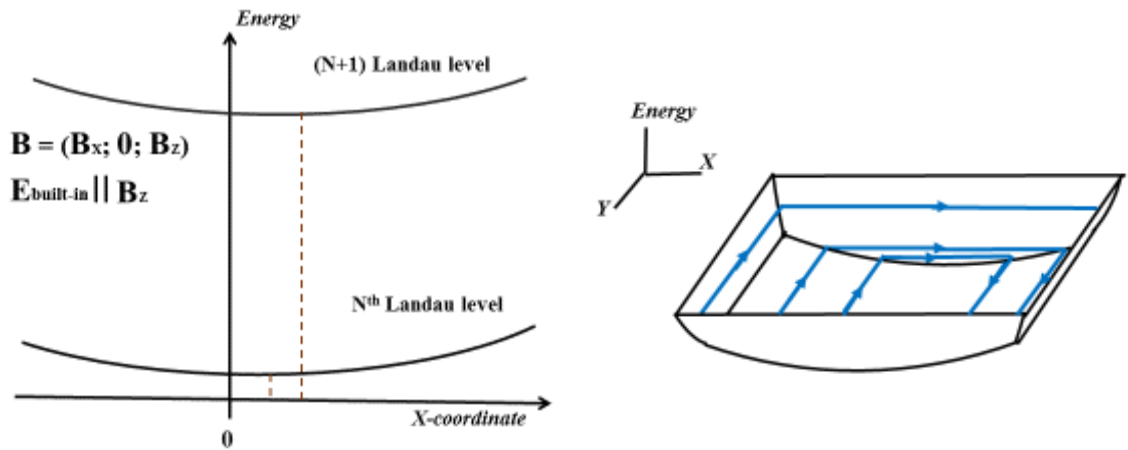

**Fig. 3**. Left panel: Calculated energy spectrum of an asymmetric IQH system in tilted quantizing magnetic field. Landau level degeneracy is lifted along the X axis. The scale is strongly exaggerated near the zero point to show a small shift of Landau bands in the X-direction.
Right panel: The energy diagram of Landau band in two dimensions. Current-carrying states are shown in blue.

At present, however, the spectrum in Fig. 3 together with the system of orbit-like quantum states is rather a suggestion because it has not been confirmed by any experiments. Thus, our chance to confirm this spectrum is to use an experimental method which has never been applied to the system of that kind but has specific features that make it promising for this purpose.

*Testing the quantum phase with macroscopic electron orbits*

To our opinion, such method does exist though it is not yet widespread. We mean the so-called intensive terahertz laser spectroscopy (for a review, see [15]). More specifically, we are going to measure terahertz-radiation-induced in-plane electric currents related to the inner system asymmetry related to a small shift of Landau bands in the *X*-direction, which ultimately is related to a non-zero vector product $B_X \times E_{\text{built-in}}$. Phenomenologically, it is the so-called photovoltaic effect (for a review, see, e.g., [16]).

In the context of our IQH-type system, the advantage of the method is, just like the magnetotransport measurements of von Klitzing, it implies the measurement of macro-currents that cannot be provided by any microscopic quantum states. The only difference is now the currents are caused not by an external electric bias but by terahertz laser excitation. A one more advantage is that by using of high-power short terahertz laser pulses we can detect a photovoltaic effect even in the case of an extremely weak in-plane asymmetry undetectable for any other methods. Finally, the energy of terahertz quanta is of the order of the energy gap between the Landau levels in the structures of type GaSb-InAs-GaSb when they are in the IQH regime whereas the expected asymmetry should manifest itself precisely under the vertical electron transitions between the Landau bands slightly shifted from each other in the *X*-direction.

Thus, if the energy spectrum in Fig. 3 is relevant, then, in presence of an asymmetry of confining potential in IQH system, the in-plane magnetic field may initiate a quantum phase transition which is accompanied by the breaking of translational symmetry in the *X*-direction and gives rise to the macroscopic quantum orbits of Halperin type in the system interior. And such transition could manifest itself under the so-called cyclotron resonance (CR) conditions when the energy of terahertz quanta is close to the gap between the Landau bands.

In the experiments to test our suggestion, we use the so-called single-quantum-well structures of type GaSb-InAs-GaSb with the well width of 15nm. The structures were grown by the method of molecular-beam epitaxy (MBE). These are the so-called semi-metallic quantum structures as the top of GaSb valence band is well above the bottom of InAs conduction band. As a result, the low-temperature electron density is as high as about $1.5 \cdot 10^{12} \text{cm}^{-2}$. To avoid the so-called electron hybridization related to the band overlapping, the InAs layer is sandwiched between two thin AlSb layers (3 nm each) because the top of AlSb valence band is below the bottom of InAs conduction band.

The low-temperature scattering time is determined by the point-like charged defects of a high concentration and as short as about 3ps. Accordingly, the electron mean free path is as short as less than 0.1μm. "Built-in" electric field is supposed to be due to the penetration of surface potential into the well because the penetration depth in such structures is known to be about 100nm whereas the surface-to-well distance is about 20nm [17].

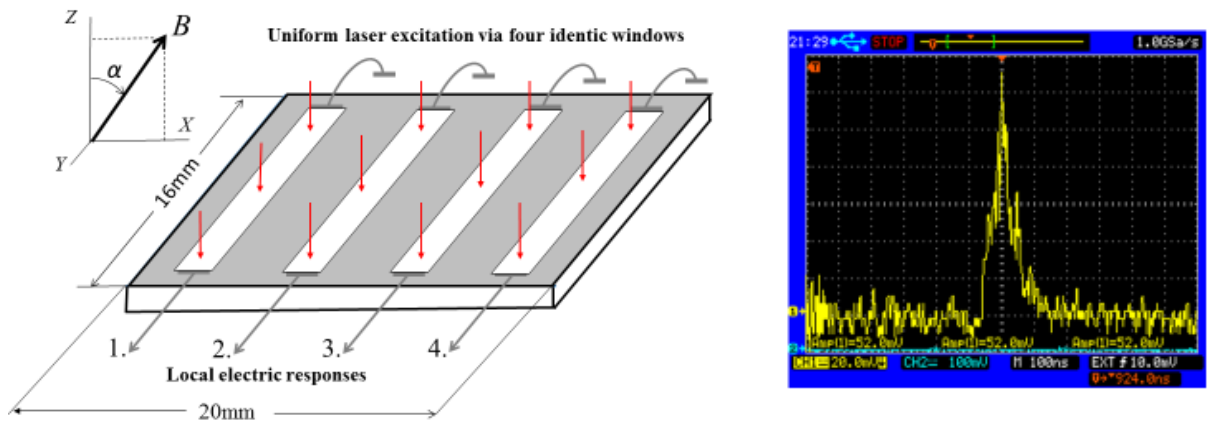

**Fig. 4.** Left panel: Sketch of the experiment with four identical windows, through which terahertz laser radiation enters an asymmetric two-dimensional structure in tilted magnetic field.
Right panel: Typical track of terahertz laser pulse (100ns/div.).

Typical sample size is 20mm in length (*X* axis) and 16mm in width (*Y* axis). It is covered with a non-transparent mask that has four identical windows shifted from each other along the *X* axis (Fig. 4, left panel). The size of each window is 2mm along the *X* axis and 12mm along the *Y* axis. The distance between the adjacent windows is 3mm. The windows are symmetrical with respect to the sample center. Each window is supplied with a pair of ohmic contacts to measure local current along the *Y* axis. The source of terahertz radiation is a pulsed ammonia laser optically pumped by $CO_2$ laser. The wavelength of terahertz radiation is 90.6μm ($\hbar\omega$ = 13.7meV), pulse duration is about 40ns, and intensity is up to 200W/cm². Typical track of terahertz laser pulse is shown in Fig. 4 (right panel). Magnetic field of up to 6T is provided by a superconducting magnet. The field can be tilted from the normal by an arbitrary angle *α* in the *XZ* plane. Sample temperature is about 2K. We measure the terahertz-radiation-induced electric responses with the kinetics similar to that of the laser pulses.

Fig. 5 (left panel) shows the dependence of local responses on the total magnetic field at high tilting angles. It is seen that here their behavior is quite trivial. They all are almost the same within the experimental error and, in the absence of quantizing magnetic field, they are simply proportional to the in-plane field. At the same time, if there is a nonzero quantizing component, the responses are strongly suppressed by the Landau quantization.

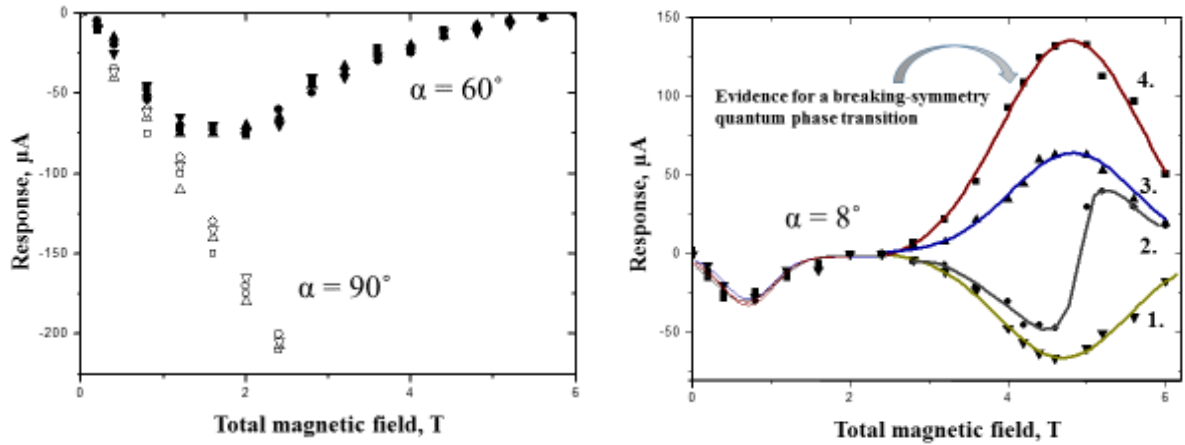

**Fig. 5.** Left panel: The dependence of photo-responses in each window on the total magnetic field at high tilting angles: blank symbols – *α* = 90˚; solid symbols – *α* = 60˚. Each symbol corresponds to its own window.
Right panel: The same dependence when the tilting angle is as low as about 8˚. Solid curves are a guide for the eyes.

But the picture changes drastically at lower tilting angles. Fig. 5 (right panel) shows the behavior of responses at α = 8˚. Initially, at low magnetic fields, the behavior is trivial again: all responses quickly disappear due to the Landau quantization. However, as we approach CR conditions (about 4.8T), responses reappear and show a resonance-like behavior with the maximum near CR point. But the most important fact is that all responses differ drastically from each other and this means that the breaking of translational symmetry does occur in accordance with our expectations.

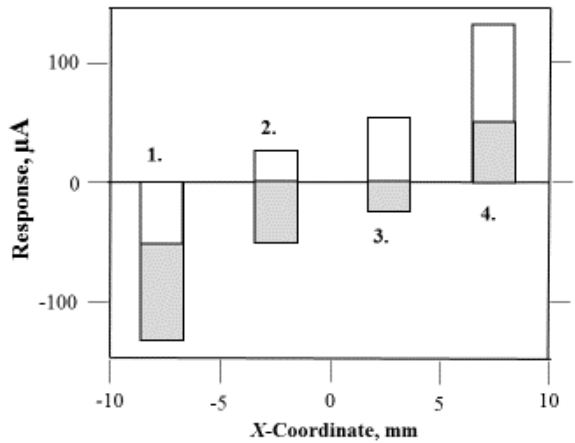

**Fig. 6.** Spatial diagram of photo-responses at the total magnetic field of 4.8T for two opposite directions of the in-plane component: empty rectangles – *α* = 8˚; solid rectangles – *α* = − 8˚.

To be sure that the effect is truly related to a macroscopic ordering of electrons, we measure responses for two opposite directions of the in-plane magnetic field when the total magnetic field is about 4.8T. The results are presented as a spatial diagram (Fig. 6). It is seen that the switching of the in-plane field does lead to a macroscopic re-ordering so that the spatial distribution of responses turns 180° around the center of the system. Thus, all experiments support the idea of a quantum phase consisted of spatially-separated macroscopic orbits and this success gives us an opportunity to realize our thought experiment and thereby to answer the question whether or not the single-electron discrete dynamics exists in nature.

*Testing the single-electron discrete dynamics*

The ultimate test is as follows. As the spatial distribution of macro-orbits is almost symmetrical with respect to the sample center, then, in accordance with the model of Fig. 2, the photo-excitation of the region No.1 would have to result in arising of photo-electrons not only in this region but also in the region No.4. Moreover, the number of such electrons would have to be *almost the same* in both regions despite an impassable wall of charged scatterers between these regions, which is *five orders thicker* than the electrons' mean free path.

In many respects, this test resembles that one in our preliminary experiments [18]. But, taking into account a fundamental character of the consequences of the test, now it will be carried out in such a way as to avoid any ambiguity. To this aim, we will compare the system behavior in two regimes – the regime of Bloch electrons and the regime of spatially separated electrons – and will do that *in situ*, i.e. only by the rotation of magnetic field in the *XZ* plane.

Moreover, now we will compare the relative number of photo-electrons in different regions. For this purpose, local responses will be used as a "counter" of photo-electrons and, to take into account that the sensitivity of such "counter" is a function of the *X*-coordinate, all responses will be normalized to the local sensitivity determined in advance when all regions are illuminated. In this case, we can make something like a map that demonstrates the spatial distribution of photo-electrons over the system under various local excitations.

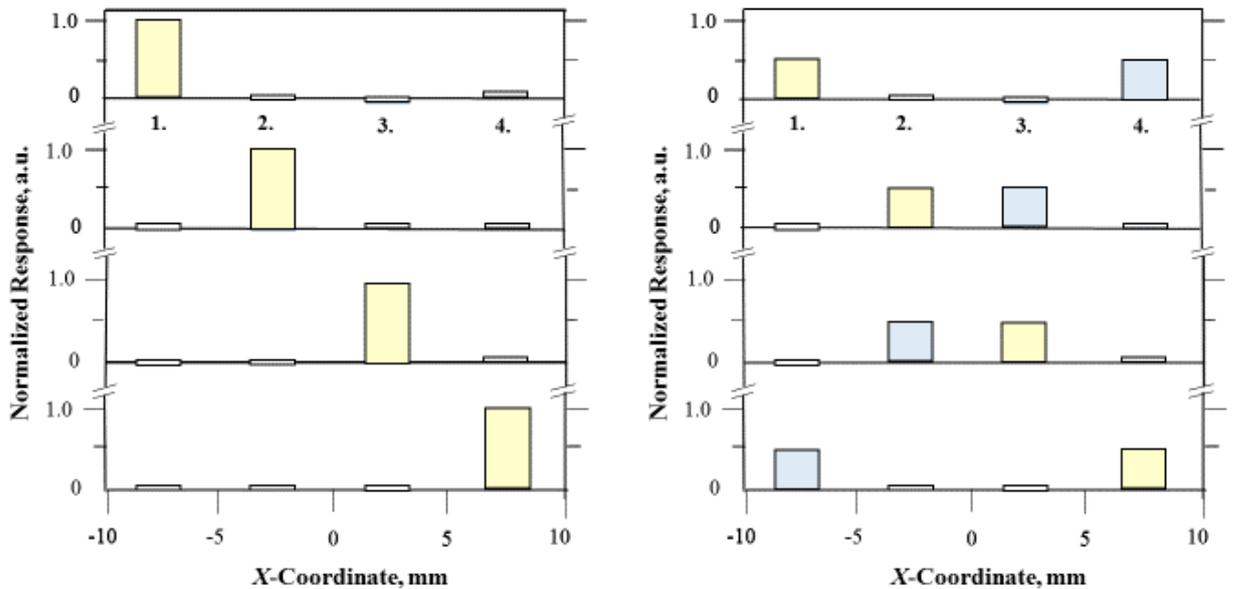

**Fig. 7.** Left panel: Spatial distribution of photo-electrons in the regime of Bloch electrons (α = 90°, B = 2.2T) when only one of four windows is open. Responses from illuminated region are shown in yellow.
Right panel: Spatial distribution of photo-electrons in the regime of the spatially-ordered electrons (α = 8°, B = 4.8T) when only one window is open. Responses from illuminated regions are shown in yellow while responses from regions in the dark are shown in blue.

Fig. 7 (left panel) shows the spatial distribution of photo-electrons in the regime of indistinguishable Bloch electrons is shown in. It is seen that here photo-electrons emerge only in illuminated regions and they are almost the same in any region. No photo-electrons are observed in the dark. This means that here the effect is strictly local at a macroscopic lengthscale.

But the picture changes drastically in the regime of spatially separated electrons (Fig. 7, right panel). Here the illumination of only the region No.1 does result in photo-electrons not only in this region but also in the region No.4 and the number of photo-electrons is indeed almost the same in both regions. Moreover, no photo-electrons are observed in the intermediate regions (No.2 and No.3) even though they are much closer to the laser spot. Similarly, the illumination of only the region No.4 gives rise to photo-electrons not only in this region but also in the region No.1 and the number of photo-electrons is again almost the same in both regions.

As we see, photo-electrons emerge in such a way as if we illuminate not one but two symmetrical regions simultaneously when the original laser beam is divided into two equal beams and a synchronized detection of the responses shows that there is no a delay between them within the time resolution (Fig. 8).

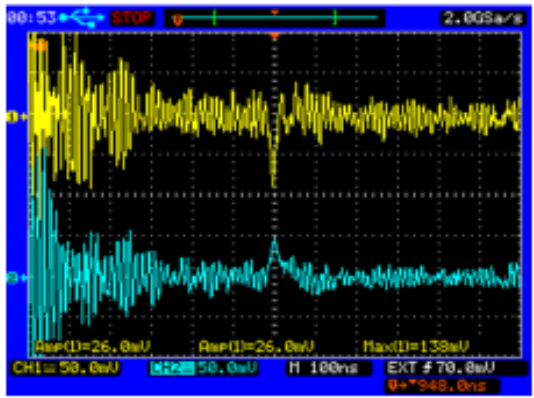

**Fig. 8.** Typical tracks (100ns/div) under the synchronized detection of photo-responses in both the illuminated region (No. 1, upper track) and the region in the dark (No. 4, lower track). The system is in the regime of macroscopic ordering ($\alpha = -8°$, $B = 4.8$T).

Our observations clearly cannot be interpreted in terms of an electron motion with a trajectory because the distance covered is five orders longer than the electron mean free path. Perhaps the only more or less reasonable interpretation in terms of a continuous motion is to assume that the excitation gives rise to some long-lived quasiparticles (something like excitons) that move quickly away from the laser spot without any scattering, and then decay at a fixed macroscopic distance giving rise to a current.

Certainly, this assumption cannot explain why only a half of quasiparticles moves away while the other half stands still. But the damning argument against this idea is that the illumination of the region No.2 gives rise to photo-electrons only in the region No.3 but not in the region No.1 despite the fact that these regions are remote from the region No.2 at the same distance. In other words, the photo-electrons "know" in which side they should "move". Through this observation, we *de facto* leave no room for any interpretation other than the single-electron spatial dynamics with macroscopic discontinuity.

## Fundamental consequences

### *Quantum absolute simultaneity and the Bohm-Hiley model of undivided universe*

As we see, experiment clearly shows that EPR nonlocality is not the only realizable type of quantum nonlocality. There is an alternative type of quantum nonlocality which is even deeper insofar as it allows nonlocal signaling. It is the single-particle discrete spatial dynamics (DSD), which ultimately is related to the fundamental indivisibility of quantum orbits where electron has no any definable position regardless of their lengthscale and spatial configuration. Mathematically, this is due to the appeal of quantum formalism to Hilbert space where any quantum orbit is a fundamentally indivisible point.

Thus, to avoid causality paradoxes, we should come back to the Bell-Popper idea and to replace the "kinematic" version of relativity by the "dynamic" version and thereby rebirth the classical insight of space and time. In fact, DSD nonlocality opens the door to a quantum concept of absolute simultaneity that now rests not on the meaningless notion of an infinite speed but on the empirically-tested DSD effect. And, in the light of this concept, we should answer the objections against this approach.

Let us start with the problem of a preferred reference frame, which currently is identified with the problem of so-called "aether". In fact, such identification is a consequence of the relativistic view of the entire universe regarded as a collection of interacting physical bodies. But there is a quantum view of the universe, which rests on the Bohm-Hiley model of undivided universe [19]. The novelty is that now we can supplement this model with the concept of quantum absolute simultaneity and thereby introduce both "true" length and "true" time without to involve the notion "aether". And then, for any physical body regarded separately from the universe, the rest of the universe appears precisely the preferred reference frame we need. In this case, the "dynamic" version of relativity surprisingly appears no less light and elegant than the "kinematic" version.

But perhaps the most formidable problem is that the purely "algorithmic" interpretation of standard QM seems incompatible with a new status of quantum theory now responsible for our view of space and time and ultimately of the entire universe. At first glance, the only way is to overcome this problem is to address the de Broglie-Bohm pilot-wave theory [20]. Currently, this theory is regarded as the only realistic version of quantum theory, which does not contradict all known experiments perhaps to the exclusion of so-called protective measurements (see, e.g., [21]). Anyway, such a solution was proposed by Bell after EPR experiments and also was discussed by us after the first experiments related to DSD.

Actually, however, this solution should be rejected. The point is the characteristic feature of the de Broglie-Bohm theory is that here electron always has a well-defined spatial position and hence a well-defined trajectory. At the same time, as we have already stressed, the macroscopic electron transport in DSD effect cannot be interpreted in terms of any trajectories (no matter Bohmian or any other) because the distance covered is five orders longer than the electron mean free path. Thus, it is the DSD effect that allows one to make the ultimate choice between the standard QM and the de Broglie-Bohm theory and this choice is clearly not in favor of the latter.

*A new approach to quantum realism*

It may seem that we have come to a deadlock because no realistic interpretation of quantum theory is currently known other than the de Broglie-Bohm theory. But this is not quite so. The point is the rejection of relativistic kinematics opens the door to a new approach of how to interpret standard QM, which ultimately rests on a deeper insight of what we call "reality".

To clarify what we mean, it first should be noted that, in the frames of relativistic kinematics, what we call "reality" is always related to the four-dimensional spacetime which is a closed structure in the sense that no more spaces can be involved into the notion "reality" because they will be out of time strictly tired precisely to conventional three-dimensional space. But if we come back to the classical view of space and time, then the time becomes again a quite separate concept. As a result, what we call "Hilbert space" regarded today as a purely mathematical concept, now can be regarded as a peculiar space related to a deeper quantum reality.

In this case, Schrödinger equation describe not the evolution of "non-existent" quantum objects in a "non-existent" quantum world (as it currently looks like) but the evolution of real quantum objects in a real quantum space. Accordingly, the wavefunction is not a pure mathematical abstraction which, for crazy reason, behaves as a real physical entity but a real physical entity indeed, which, however, is related not to the observable three-dimensional reality but to a deeper unobservable quantum reality.

What is really important here is the both Hilbert space and Euclidean space have the same notion of time and that quantum objects can be transferred from the former into the latter. This transfer is related to what we call "measurement" known to play a key role in the standard QM. But now this role becomes more evident. Measurement is the gateway through which quantum objects could, in accordance with the Born rule, enter the observable reality interrupting thus their evolution in Hilbert space.

In fact, through the concept of a deeper reality we revive the Einstein's concept of objective reality that currently seems incompatible with QM. And, in the vein of this concept, in our insight of measurement we do not follow the most popular Heisenberg-von Newman's line where measurement always implies a real observer and therefore has a subjective character [22, 23]. Instead, we follow the Landau-Lifshitz's line where measurement looks rather as a physical process when quantum object becomes a part of a macroscopic body (say, a device) and is forced, for a time, to become a part of observable reality and hence to take only one of its eigenstates [1]. In this case, the presence of an observer is optional and therefore such process may occur at any time as well as in any point of universe.

Finally, the concept of a deeper reality allows one to resolve many quantum paradoxes. Let us take, for example, the famous paradox of Schrödinger cat. Here a radioactive atom, as a quantum object, belongs to a deeper reality where it may well be in a superposition of two states: excited and non-excited. At the same time, any superposed state is a prerogative of deeper reality and this state by no means can be transmitted to a macroscopic body (a cat) belonged to the observable reality. Therefore, for the cat, such a situation is not more than a sort of "heads and tails" game but where the probability of unwanted result is an increasing function of time.

A separate question in the context of our observations is what should we do with the relativistic theory of gravity, i.e. with the GR. The answer is that, although GR remains the most popular theory of gravity, today we know a number of alternative approaches (for a review, see, [24]). Moreover, some of these approaches make an attempt to involve quantum principles in the solution of the problem. It is the so-called quantum gravity. And since, in the minds of many physicists, gravity inextricably linked with the theory of relativity, quantum gravity is often regarded as a platform where the long-awaited unification of quantum and relativistic theories will eventually take place as a symbol of harmony of our insights of physical world.

Leaving aside the question of fruitfulness of the attempts to involve quantum theory into the problem of gravity, here we only note that, in terms of the concept of a deeper reality, a harmony of physical world should not necessarily be viewed as a direct unification of quantum and relativistic theories. Actually, this view is based on the belief that all physical theories have the same subject – the observable reality. But insofar as the subject of QM is an unobservable reality that obeys its own laws, this theory is connected with the observable reality only indirectly, through the process of measurement. In this case, our view of a harmony of our insights of physical world should be more flexible in the sense that we should not suppose that the observable world must obey some universal physical laws. Rather, we should require that the testable predictions of all physical theories should not contradict our observations.

*Final remark*

Finally, there is also a subjective reason why it is hard to accept the Bell-Popper idea. The point is, for more than a hundred years, relativistic kinematics reigned supreme in physics and many generations of physicists were brought up in the spirit of these ideas literally from their childhood. As a result, the concept of relativistic spacetime is so firmly established in scientific usage that it is extremely hard to abandon it. However, as we know from the history of physics, the prevailing concept of space and time is always determined by what we currently know about the spatial dynamics of physical bodies. Therefore, any new fundamental physical theory may change this concept in a revolutionary way. This was happened at the beginning of the last century, when the classical concept of space and time was replaced by the relativistic concept when a deeper relativistic dynamics was confirmed experimentally. Certainly, at that time, it was extremely hard to see the next "revolutionist" in the just birth QM. Only after a hundred years, the fantastic predictive power of QM together with its extremely successful expansion into the macro-world is forced us to bring our current ideas about space and time into line with quantum laws. And as a great benefit of this step, it would help us not only to reconcile QM and relativity but also to come back to a realistic view of physics, which, despite everything, Einstein advocated throughout all his life.

## Acknowledgements

The Author is grateful to Prof. Sergey Ivanov (Ioffe Institute) for the MBE samples as well as to Prof. Raymond Chiao (UC at Merced) for useful comments on the experiment.